\documentclass[preprint2]{emulateapj}
\usepackage{natbib}
\shorttitle{Nonmetastable Ammonia Masers in NGC 7538}
\shortauthors{Hoffman \& Joyce}

\begin{document}

\title{New Maser Emission from Nonmetastable Ammonia in NGC 7538. IV. Coincident Masers in Adjacent States of Para-ammonia}

\author{Ian M.\ Hoffman}
\affil{Physics Department, Wittenberg University, Springfield, OH 45501}
\email{ihoffman@wittenberg.edu}

\and

\author{Spenser A.\ Joyce}
\affil{Physics Department, Wittenberg University, Springfield, OH 45501}

\begin{abstract}
We present the first detection of para-ammonia masers in NGC~7538: multiple epochs of observation of the $^{14}$NH$_3$ $(J,K) = (10,8)$ and (9,8) lines.
We detect both thermal absorption and nonthermal emission in the (10,8) and (9,8) transitions and the absence of a maser in the (11,8) transition.
The (9,8) maser is observed to increase in intensity by 40\% over six months.
Using interferometric observations with a synthesized beam of $0\farcs25$, we find that the (10,8) and (9,8) masers originate at the same sky position near IRS~1.
With strong evidence that the (10,8) and (9,8) masers arise in the same volume, we discuss the application of pumping models for the simultaneous excitation of nonmetastable $(J > K)$ para-ammonia states having the same value of $K$ and consecutive values of $J$.
We also present detections of thermal absorption in rotational states ranging in energy from $E/k_B \sim 200$~K to 2000~K, and several non-detections in higher-energy states.
In particular, we describe the populations in eight adjacent rotational states with $K=6$, including two maser transitions, along with the implications for ortho-ammonia pumping models.
An existing torus model for molecular gas in the environment of IRS~1 has been applied to the masers; a variety of maser species are shown to agree with the model.
Historical and new interferometric observations of $^{15}$NH$_3$ (3,3) masers in the region indicate a precession of the rotating torus at a rate comparable to continuum-emission-based models of the region.
We discuss the general necessity of interferometric observations for diagnosing the excitation state of the masers and for determining the geometry of the molecular environment.
\end{abstract}


\section{Introduction}

Six Galactic star-forming regions are known to exhibit nonmetastable ($J > K$) maser emission from ammonia (NH$_3$).
Masers from both ortho-ammonia ($K = 3n; n = 0, 1, 2, \ldots$) and para-ammonia ($K \ne 3n$) are known.
NGC~7538 has been observed to exhibit $(J,K) = (9,6)$ and (10,6) maser emission from $^{14}$NH$_3$ (Madden et al.\ 1986; Hoffman 2012, hereafter Paper~III) and (4,3) and (3,3) maser emission from $^{15}$NH$_3$ (e.g., Schilke et al.\ 1991).
In the Galaxy overall, four of the known maser regions were discovered by Madden et al.\ (1986) in a survey of 17 objects for (9,6) and (6,3) emission with a sensitivity of $\approx$100~mJy.
The other two regions were discovered by Walsh et al.\ (2007, 2011) in a survey of $>500$ objects for (11,9) and (8,6) emission with $\approx$2-Jy sensitivity.

An open question in the study of nonmetastable ammonia is the interrelationship among the masering lines, especially those with the same azimuthal angular-momentum quantum number $K$.
The relationship among states that are ``rungs'' of total angular momentum $J$ on a ``ladder'' of common value $K$ is determined by the excitation mechanism.
The masers occur in the familiar $\sim 20$-GHz symmetric-antisymmetric inversion doublets of the rotational states in the vibrational ground state.
Because the upper halves of every doublet in a ladder have the same parity, rotational transitions within a ladder could only ever overpopulate alternating rungs' doublets.
It was recognized by Mauersberger, Henkel, and Wilson (1987, 1988) that transitions from ground-state rungs through an intermediate vibrationally-excited rung would allow the necessary overpopulation of adjacent ground-state doublets.
Brown and Cragg (1991) and Schilke et al.\ (1991) developed quantitative pumping models in which a nearby source of infrared radiation provides the necessary vibrational excitation ($\approx 1400$~K above the ground state) and subsequent relaxation in order to produce masering in adjacent rungs in the vibrational ground state (see also Henkel et al.\ 2013, and references therein).

Different research groups are now solving the two-decade-old observational problem of applying this pump model by locating ammonia volumes that support maser emission from several adjacent rungs.
In Paper~III, using interferometric imaging, we reported the physical association of some -- but not all -- of the $^{14}$NH$_3$ (10,6) and (9,6) masers with common velocities in NGC~7538.
Henkel et al.\ (2013), using common temporal velocity variability among several single-dish observations, have successfully associated (6,6), (7,6), \& (8,6) masers and (9,9), (10,9), \& (11,9) masers with the same volume of gas in W51 IRS~2.
In contrast to these ortho-ammonia results, no two volumes of adjacent-state para-ammonia masers have been correlated using either velocity variability or interferometric imaging.

Prior to Papers~I (Hoffman \& Kim 2011) and III in the current series, interferometric observations of nonmetastable ammonia masers were rare and have never included more than one rung in a ladder; the four previous studies are
(1) a 3\farcs1-resolution observation of (9,8) in W51 by Wilson, Johnston, and Henkel (1990),
(2) a 0\farcs02-resolution VLBI observation of (9,6) in W51 by Pratap et al.\ (1990), 
(3) a 0\farcs2-resolution observation of (9,8) in W51 by Gaume et al.\ (1993), and
(4) a 1\farcs0-resolution observation of (11,9) and (8,6) in NGC~6334I by Walsh et al.\ (2007).
With the current paper, we seek to provide interferometric constraints for the first time to the pumping of nonmetastable para-ammonia masers.

As described in Paper~III, there is a lack of consensus in the literature concerning the importance of the levels in the $K=0$ ladder to the excitation of ortho-ammonia masers with $K=3$ and $K=6$, compared with vibrational promotion.
Since the excitation of para-ammonia cannot involve the $K=0$ states, a study of para-ammonia masers in adjacent rungs of the same ladder is a uniquely valuable diagnostic of vibrational pumping schemes.
In order to provide a robust sampling of three or more adjacent rungs of nonmetastable ammonia suitable for constraining prevailing excitation models, we have observed several adjacent ammonia transitions in NGC~7538 -- in particular the (11,8), (10,8), and (9,8) transitions of para-$^{14}$NH$_3$ and the lowest eight $K=6$ levels of ortho-$^{14}$NH$_3$ -- using the Robert C.\ Byrd Green Bank Telescope (GBT) and Karl G.\ Jansky Very Large Array (VLA) of the NRAO.\footnote{The National Radio Astronomy Observatory is a facility of the National Science Foundation operated under cooperative agreement by Associated Universities, Inc.}

\section{Observations and Results}

\subsection{GBT Observations of $^{14}$NH$_3$ and $^{15}$NH$_3$\label{GBT}}

Using the GBT on four different dates spanning 15 months, we observed 29 different transitions of $^{14}$NH$_3$ and two different transitions of $^{15}$NH$_3$.
The observational parameters for $^{14}$NH$_3$ are summarized in Table~\ref{absorption}.
In all cases, the spectral resolution was at least as fine as $0.1\,{\rm km}\,{\rm s}^{-1}$, suitable for resolving the widths of any maser lines ($\Delta{v} \approx 0.7\,{\rm km}\,{\rm s}^{-1}$).
All of the GBT observations are averages of dual circular polarizations.
The flux calibrator used for all observations was 3C48, except for 3C295 in 2012 March.
In all cases, J2322+5056 was used for pointing and secondary calibration.
The GBT observations were reduced using GBTIDL\footnote{The GBT Interactive Data Language is documented at http://gbtidl.nrao.edu}, including fitting and subtraction of the baseline of continuum emission from NGC~7538.

\begin{figure}
\centering
\includegraphics[angle=000,scale=0.40]{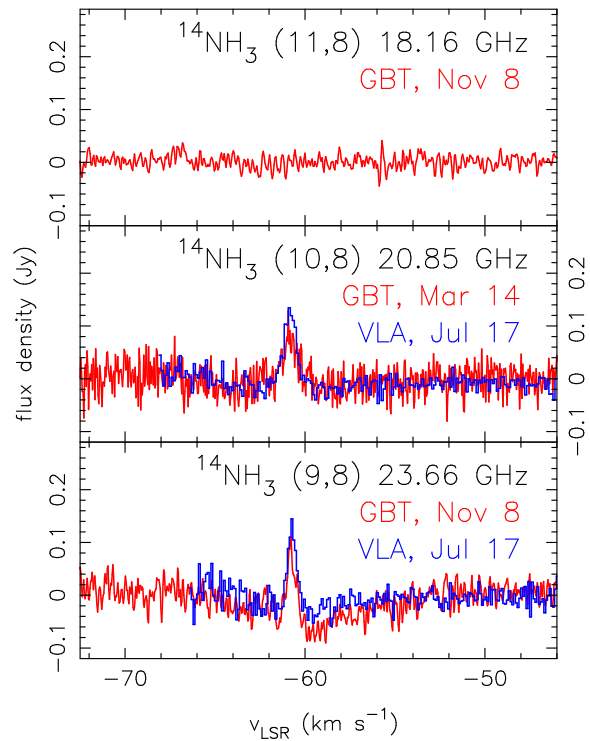}
\caption{
Spectra from the lowest three nonmetastable rungs of the $K=8$ ladder of $^{14}$NH$_3$ in NGC~7538.
The red curves are GBT observations and the blue histograms are VLA observations.
The absorption properties are summarized in Table~\ref{absorption} and the emission properties are summarized in Table~\ref{emission}.
As shown in Figure~\ref{sky}, the (10,8) and (9,8) emission arises at the same location in IRS~1.
There is no emission or absorption detected for (11,8), as discussed in Section~\ref{pump}.
	All observation dates are 2012.
\label{K=8}}
\end{figure}

\begin{figure}
\centering
\includegraphics[angle=000,scale=0.40]{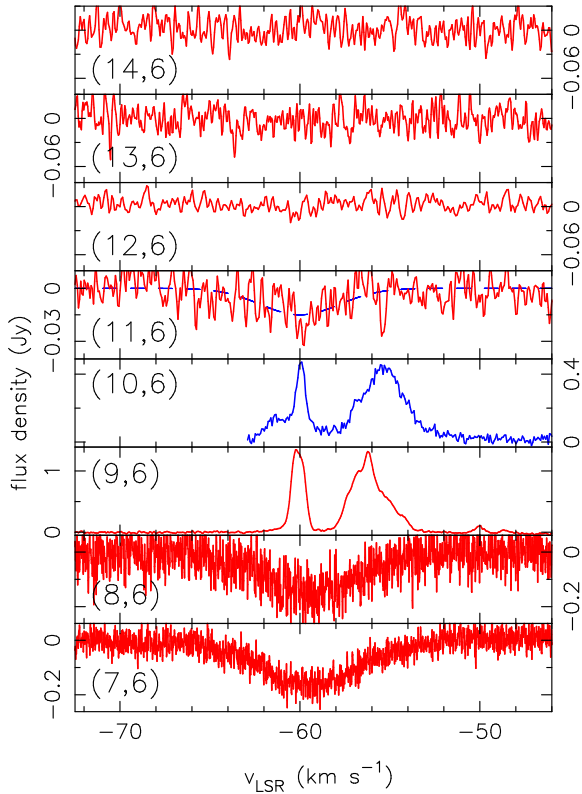}
\caption{
	Spectra from the GBT observations of the lowest eight nonmetastable $K=6$ ladder transitions of $^{14}$NH$_3$ in NGC~7538.
	The absorption parameters are summarized in Table~\ref{absorption}.
	The dashed line shows the fit to (11,6); no other fits are shown.
	The maser emission in (9,6) and (10,6) is described in Paper~III.
	The implications of the thermal detections to the maser pumping is discussed in Section~\ref{pump}.
\label{ladder}}
\end{figure}

\begin{figure}
\centering
\includegraphics[angle=000,scale=0.40]{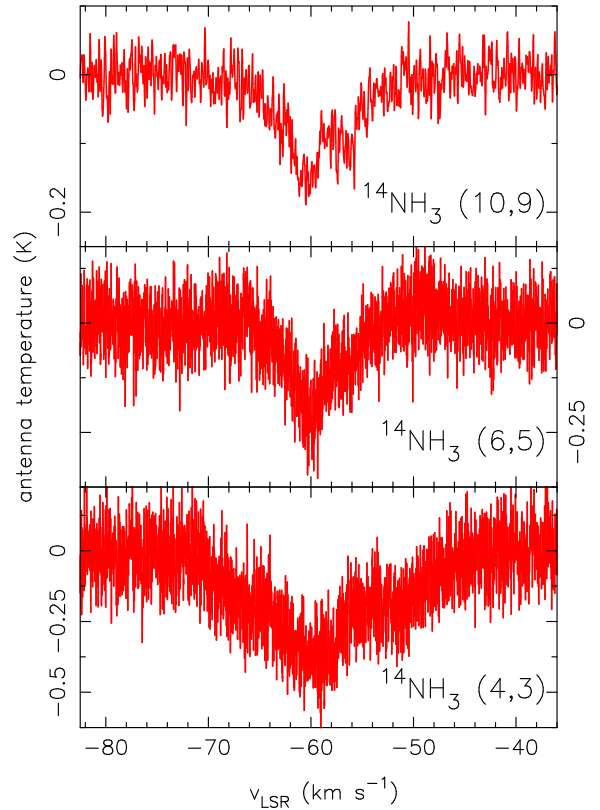}
\caption{
	A selection of spectra from the GBT observations of $^{14}$NH$_3$ in NGC~7538.
	These spectra show significant velocity structure, as summarized in Table~\ref{absorption}.
	Note that although the $^{14}$NH$_3$ (4,3) shown here is in absorption, the $^{15}$NH$_3$ (4,3) shown in Figure~\ref{15NH3} is in emission.
\label{14NH3}}
\end{figure}

Fourteen of the $^{14}$NH$_3$ transitions were observed in absorption.
Among these are the first astronomical detections of (12,9) and (11,6).
At $E/k_B = 1789$~K, the (12,9) line is the highest-energy nonmetastable rotational state yet detected in the vibrational ground state.
The fitted parameters of the absorption profiles are summarized in Table~\ref{absorption} along with the upper limits of the non-detections.
The red lines in Figures~\ref{K=8}, \ref{ladder}, and \ref{14NH3} are selected GBT spectra.

\begin{figure}[t]
\centering
\includegraphics[angle=270,scale=0.33]{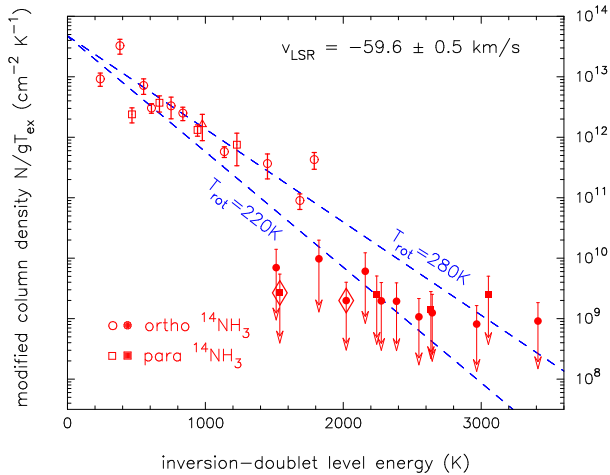}
\caption{
	Boltzmann plot of the GBT observations of nonmetastable ammonia in NGC~7538.
	The open symbols are detections, the filled symbols are upper limits from non-detections, the circles are transitions of ortho-ammonia, and the squares are transitions of para-ammonia.
	The triangle symbol is the measurement of $(8,3)$ from Paper~III.
	Masers have been excluded since they deviate from thermal behavior.
	The thermal absorptions presented here are observed with a central velocity in the range $v_{\rm LSR} = -59.6 \pm 0.5$~km/s.
	The horizontal axis is $E_{\rm rung}$ in Table~\ref{absorption}.
	The vertical axis is the column density modified by normalizing by the statistical weight of the level, $g$, and the excitation temperature, $T_{ex}$, as discussed in Section~\ref{GBT}.
	The slope of a best-fit straight line is indicative of the rotational temperature, $T_{\rm rot}$, of the thermal population.
	The dashed lines represent the expected line strengths for temperatures of $T_{\rm rot} = 280$~K and 220~K.
	The diamond symbols highlight the non-detections of $(11,8)$ and $(12,6)$ discussed in Section~\ref{pump}.
\label{boltz}}
\end{figure}

The relative amplitude of the various absorption lines are well fit by a thermal interpretation of the populations in the levels.
Since these radiofrequency transitions of ammonia are low-energy symmetric-antisymmetric inversions of a relatively high-energy $(J,K)$ rotational state, the expected absorbing population is calculated from the energy of the rotational state, listed as $E_{\rm rung}$ in Table~\ref{absorption}.
Following, for example, Schilke et al.\ (1991), we have made a Boltzmann plot of the GBT observations, shown in Figure~\ref{boltz}.
For the 14 detections, the best-fit temperature to the rotational populations is $T_{\rm rot}=280$~K.
A fit that incorporates the non-detections yields a lower value of $T_{\rm rot}=220$~K.
Both fits are shown in the figure.
Our value for $T_{\rm rot}$ in NGC~7538 is in reasonable agreement with previous measurements using ammonia: 170~K (Wilson et al.\ 1983), 220~K (Mauersberger et al.\ 1986b), 150~K (Schilke et al.\ 1991).
Since a single temperature fits well all of the data, we assume that the ammonia in the cloud at $v_{\rm LSR}=-59.6$~km/s is well thermalized and internally equilibrated (there are no multimodal distributions or gradients).

In Figure~\ref{K=8} are shown the spectra from the three adjacent rungs in the $K=8$ ladder.
For the first time in NGC~7538, we have detected emission for the (10,8) and (9,8) lines that we argue is maser emission.
The emission properties are summarized in Table~\ref{emission}.
The spectra were Hanning-smoothed offline resulting in an rms noise of $\approx 20$~mJy in a 0.04-km/s channel.
Although inter-epoch and inter-instrument comparisons are limited by $\approx$20\% calibration uncertainties in flux scale (see also Henkel et al.\ 2013), we assume that the level of continuum emission from IRS~1 did not change over the six months between GBT observations.
In comparing line strength with continuum level, we find that the (9,8) maser increased in intensity by $\approx$40\% from March to November 2012.
There is only a single epoch of GBT observation of the (10,8) line and so we have no variability constraints for (10,8).
We do not detect emission or absorption in the (11,8) spectrum.

\begin{figure}[b]
\centering
\includegraphics[angle=000,scale=0.30]{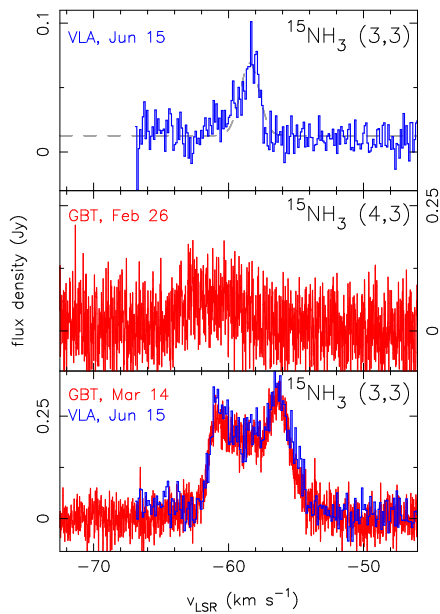}
\caption{
	Spectra of metastable and nonmetastable $^{15}$NH$_3$ in NGC~7538.
	The red curves are GBT observations and the blue histograms are VLA observations.
	The VLA spectrum in the bottom panel has been the spatially integrated with a 5-arcsecond convolving beam.
	The emission arises in IRS~1, as shown in Figure~\ref{kntr}.
	The top panel shows the spectrum of the new maser emission indicated with the box in Figure~\ref{gradient} at $\Delta\alpha=-0\farcs126(8)$, $\Delta\delta=-0\farcs45(1)$.
	As described in Section~\ref{VLA}, the synthesized beam for the top panel is $0\farcs29 \times 0\farcs23$.
	The dashed line in the top panel is the fitted spectral profile of the new maser with $S=0.56(4)$~Jy, $\Delta{v_{\rm FWHM}}=1.7(1)$~km/s, and $v_{\rm LSR}=-58.43(6)$~km/s.
	The (4,3) ``millimaser'' is discussed in Section~\ref{GBT}.
	All observation dates are 2012.
\label{15NH3}}
\end{figure}

In Figure~\ref{15NH3} are shown (in red) the GBT spectra of the (3,3) and (4,3) states of $^{15}$NH$_3$.
Both lines are observed in emission, as previously reported by Gaume et al.\ (1991) and by Schilke, Walmsley, and Mauersberger (1991).
The (4,3) ``millimaser'' emission previously described by Schilke et al.\ is not analyzed further in the current paper.
Nevertheless, we note that we used a rest frequency of 21.637730~GHz, resulting in a shifted velocity for our profile compared to theirs (see their discussion concerning their rest frequency).
The (4,3) amplitudes of the current spectrum and of the 1990 spectrum are not significantly different.

\subsection{VLA Observations of $^{14}$NH$_3$ and $^{15}$NH$_3$\label{VLA}}

On 2012 July 17, we observed the (10,8) and (9,8) lines of $^{14}$NH$_3$ using the VLA.
The VLA observations employed one circular polarization in two separate bands during the same pointing.
The sources used for flux and bandpass calibration were 3C48 and 3C147, with J2322+5056 used for pointing and secondary calibration.

In `B' configuration, the resulting synthesized beam was $0\farcs30 \times 0\farcs22$ at a position angle of $-8\degr$.
The rms noise in a single 0.1-km/s channel image is $13\,{\rm mJy}\,{\rm beam}^{-1}$, consistent with instrumental expectations.
The continuum emission from NGC~7538 was imaged and subtracted in the $u,v$ plane as part of standard reduction using AIPS\footnote{The Astronomical Image Processing System is documented at http://www.aips.nrao.edu/index.shtml}.
The spectra at the peak image location are shown overlaid in Figure~\ref{K=8}.
The deconvolved sizes of the maser images are listed in Table~\ref{emission} along with the corresponding brightness temperatures of the emission.
The masers have angular sizes of $\approx100$~mas ($\approx 300$~AU at a distance of 2.65~kpc, Moscadelli et al.\ 2009).

\begin{figure}
\centering
\includegraphics[angle=270,scale=0.45]{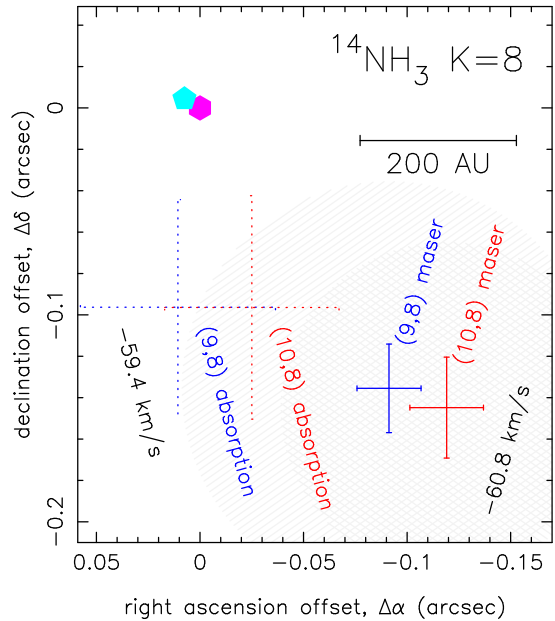}
\caption{
The sky positions of the (10,8) and (9,8) image features from the VLA observations of NGC~7538 IRS~1.
The observations have a synthesized beam of $0\farcs30 \times 0\farcs22$ at a position angle of $-8\degr$.
The maser emission and absorption are observed at different velocities, as indicated.
The solid crosses at the lower right of the panel are the positions of the masers.
The size of the crosses represent the 1-$\sigma$ formal fitting error to the position of the emission.
The light gray hatched regions represent the deconvolved sizes of the masers, as listed in Table~\ref{emission} (lower-left to upper-right hatching for (9,8), lower-right to upper-left for (10,8)).
Crossed hatching indicates coincident, overlapping emission from the (9,8) and (10,8) masers.
The dotted crosses at the left of the panel are the locations of the absorption from the two lines.
As discussed in Sections~\ref{VLA} and \ref{registration}, all positions are displayed as offsets from the fitted location of the peak of the continuum emission in IRS~1 at $\alpha_{\rm J2000} = 23\, 13\, 45.37(1)$, $\delta_{\rm J2000} = 61\, 28\, 10.4(1)$. 
The filled pentagon and hexagon are the fitted locations of the peaks of 20.85- and 23.66-GHz continuum emission from IRS~1, respectively.
The absorption for (8,3) observed interferometrically (with a $1\farcs9$ beam, see Section~\ref{VLA}, Table~\ref{absorption}, and Paper~III) is not discernibly different in position than the peak of 16.4-GHz continuum emission, and is not plotted.
\label{sky}}
\end{figure}

The precision of the absolute position of the images in either of the two bands is limited to $\approx 200$~mas by interferometer phase solutions (see further discussion in Sec.~\ref{registration}).
We find that the fitted peaks of continuum emission from IRS~1 at 20.85~GHz and 23.66~GHz differ in absolute position by 20~mas and so are assumed to be coincident.
The positions of the masers relative to the peak of continuum emission are shown in Figure~\ref{sky}.
The difference in positions of the (10,8) and (9,8) masers is much less than their angular size.
Furthermore, in velocity, the difference between the centers of the (10,8) and (9,8) lines is much less than the line widths.
Therefore, the (10,8) and (9,8) masers are assumed to be coincident in position and velocity in the following discussion (see also \S~\ref{pump} for further arguments).
As shown in Figure~\ref{sky}, the absorption is offset from the masers in both velocity and position.

On 2012 June 15, simultaneous with the $^{14}$NH$_3$ (9,6) observations described in Paper~III, we observed the (3,3) line of $^{15}$NH$_3$ at $\nu_{\rm rest} = 22.789421$~GHz, employing all 27 antennas of the VLA.
In `B' Configuration, the relative separations between the antennas ranged from 0.3 to 11.1~km, resulting in a synthesized beam of $0\farcs29 \times 0\farcs23$ at a position angle of $-17\degr$; the array was not sensitive to image features larger than approximately 10~arcseconds.
The flux, phase, and bandpass calibrators used were 3C48, 2322+509, and 3C147 with approximately 45 minutes of integration on the IRS~1 target.
We recorded a single circular polarization and a 2-MHz bandwidth centered in frequency on $v_{\rm LSR}=-53.8\,{\rm km}\,{\rm s}^{-1}$ divided into 256 spectral channels resulting in a velocity resolution of approximately $0.1\,{\rm km}\,{\rm s}^{-1}$ and a total velocity coverage of approximately $26\,{\rm km}\,{\rm s}^{-1}$.
The data were reduced using AIPS.
The {\it rms} background noise in a channel image is $10\,{\rm mJy}\,{\rm beam}^{-1}$, consistent with instrumental expectations.

\begin{figure}[t]
\centering
\includegraphics[angle=270,scale=0.67]{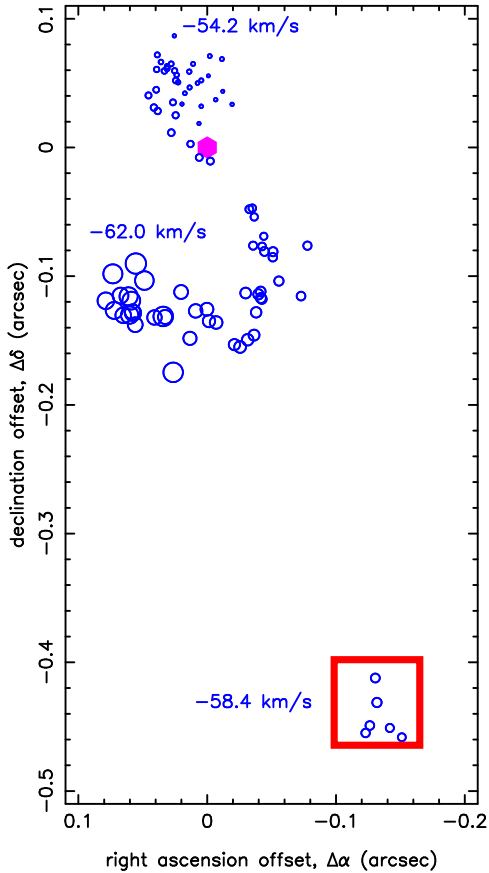}
\caption{
	The sky positions of emission peaks in the velocity channel images from the 2012 VLA observations of the $^{15}$NH$_3$ (3,3) masers in NGC~7538 IRS~1.
	The size of the circle markers represents the velocity of the feature, with larger circles representing more negative velocities.
	The velocities at the extremes of the apparent trend are labeled.
	A model for the trend is discussed in Section~\ref{model}.
	The spatially-integrated emission profile of the entire emission region is shown in Figure~\ref{15NH3}.
	Also shown in Figure~\ref{15NH3} is the spectrum at the location of the previously undetected emission indicated by the box.
	As discussed in Section~\ref{registration}, the position offsets are measured from the fitted location of the peak of 22.8-GHz continuum emission at $\alpha_{\rm J2000} = 23\, 13\, 45.37(1)$, $\delta_{\rm J2000} = 61\, 28\, 10.4(1)$.
	The continuum emission in this region is shown in Figure~\ref{kntr}.
\label{gradient}}
\end{figure}

\begin{figure}[b]
\centering
\includegraphics[angle=000,scale=0.28]{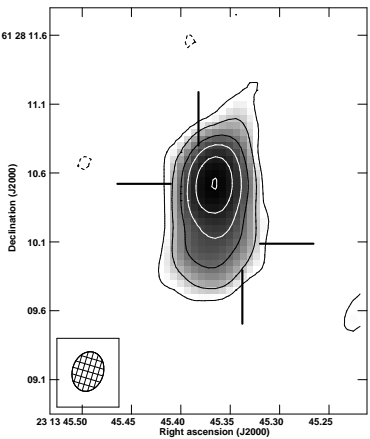}
\caption{
	VLA image of the 22.8-GHz continuum emission from NGC~7538 IRS~1.
	The beam, plotted at the lower left, is $0\farcs29 \times 0\farcs23$ at a position angle of $-17\deg$.
	The contour levels are $-$3, 3, 9, 27, 54, 108, and 192 times the image rms noise level of 900 $\mu$Jy/beam.
	The fitted peak of the emission is the offset origin in Figure~\ref{gradient}.
	The lines are the extended edges of Figure~\ref{gradient}.
\label{kntr}}
\end{figure}

All of the channel images in the range $-62.0\,{\rm km/s} < v_{\rm LSR} < -53.8\,{\rm km/s}$ contain one to three locations of $^{15}$NH$_3$ (3,3) line emission, summarized graphically in Figure~\ref{gradient}.
As noted previously by Johnston et al.\ (1989) and Gaume et al.\ (1991), the velocity trend across the sky does not follow a simple, linear gradient or direction.
We discuss the velocity structure in Section~\ref{model}.
The spatially-integrated emission observed with the VLA is in good agreement with the GBT observations of 2012 March, as shown overlaid in Figure~\ref{15NH3}.
Also shown in Figures~\ref{15NH3} and \ref{gradient} are the emission profile and location of a new source of $^{15}$NH$_3$ (3,3) maser emission at $v_{\rm LSR}=-58.4$~km/s, significantly offset in position from all previously reported locations, although still correlated with IRS~1.
Figure~\ref{kntr} is an image of the 22.8-GHz continuum emission from IRS~1, showing the boundaries of Figure~\ref{gradient}.
The integrated flux density in the continuum continuum image is 380~mJy, in good agreement with 370~mJy reported by Sandell et al.\ (2009).

On 2012 January 31, during the same observation described in Paper~III for the VLA `C' Configuration observations of $^{14}$NH$_3$ (10,6), we detected absorption of $^{14}$NH$_3$ (8,3).
This is the first astronomical detection of this transition, is included in Table~\ref{absorption}, and is included in Figure~\ref{boltz} as a triangle symbol.

\section{Discussion\label{disc}}

\subsection{Maser Pumping\label{pump}}

It is generally expected that the overpopulation of a nonmetastable ammonia inversion doublet (that is, of the upper of the two symmetric-antisymmetric states within the rung) is achieved through parity selection of the levels because the two states in the doublet are so close in energy compared to the separation among rungs.
In order to achieve the necessary parity changes, pumping models for both astrophysical masers (for example, Mauersberger, Henkel, and Wilson 1988) and for laboratory masers (for example, Willey et al.\ 1995) include an intermediate state outside of the vibrational-ground-state ladder in which the maser occurs.
For ortho-ammonia, this intermediate state may be in the vibrational-ground-state $K=0$ ladder (see discussion in Paper~III) or may be a radiative vibrational excitation with the same value of $K$: $(J,K,v=0) \to (J,K,v=1)$.
For para-ammonia, the $K=0$ states are inaccessible and so the only remaining modeled option is vibrational promotion.
If the excitation process lacks a parity-changing intermediate state, only the alternating rungs $(J,K)$ and $(J+2,K)$ could support masers, with the intervening $(J+1,K)$ rung instead being underpopulated (Brown and Cragg, 1991).

We report coincident maser emission from consecutive rungs -- $(J,K)=(9,8)$ and $(J+1,K)=(10,8)$ -- with no detectable emission in $(J+2,K)=(11,8)$.
It is possible that rather than both masers arising in the same emitting volume, there are separate (10,8) and (9,8) volumes displaced along the line of sight.
A similar displacement of two volumes is suggested by Schilke et al.\ (1991) for the collisional pumping of $^{15}$NH$_3$ (4,3) and (3,3) masers in IRS~1.
However, the occurrence of two different nonmetastable (10,8) and (9,8) masers does not require two different densities as do the $^{15}$NH$_3$ masers, nor would separate $K=8$ volumes necessarily agree with the torus model discussed in Section~\ref{model}.
We conclude that the (10,8) and (9,8) masers are physically associated with each other and are indicative of a common excitation mechanism.
This result is strong evidence for vibrational pumping of the $K=8$ maser in NGC~7538 IRS~1.

The non-detection of the (11,8) transition is highlighted with a diamond in Figure~\ref{boltz}.
It is noteworthy that the (11,8) upper limit is anomalously below the expected threshold for detection based on the fitted $T_{\rm rot}$ for the rest of the population.
The population in the (11,8) state is significant for the interpretation of the (10,8) and (9,8) masers for three reasons:
(1) alternating-rung processes described by Brown and Cragg (1991) link (11,8) to the maser in (9,8),
(2) adjacent rung-processes, such as parity selection, link (11,8) to the maser in (10,8), and
(3) the pumping mechanism (perhaps vibrational promotion) of ground-state $K=8$ levels apparently has an upper limit that does not extend above $J=10$.
For interpretation (1), we find no evidence that the alternate-rung processes (simple rotational excitation conserving intra-ladder parity) are important since the (9,8) maser is not accompanied by an (11,8) maser.
For interpretation (2), the observed underpopulation of (11,8) is expected based on a simple rotational excitation of the (10,8) maser.
Of course, the other rung adjacent to (10,8) is not underpopulated, but rather is the (9,8) maser.
Since (1) and (2) cannot simultaneously be true for rotational excitation, we conclude that (3) vibrational promotion is responsible for both the (10,8) and (9,8) masers, but is limited to rungs below (11,8) ($E/k_b\approx1500$~K).
This mechanism does not explain the observed underpopulation of (11,8), especially if the absorbing cloud at $v_{\rm LSR}=-59.6$~km/s is physically distinct from the maser-bearing cloud at $v_{\rm LSR}=-60.8$~km/s.

Ortho-ammonia observations can provide additional context about the uppermost ladder rung for which maser excitation is exhibited.
Figure~\ref{ladder} shows a complete sampling of the lowest eight levels of the $K=6$ ladder.
Above the masering (10,6) and (9,6) rungs, the (11,6) line is detected in thermal absorption in good agreement with the Boltzmann populations presented in Figure~\ref{boltz}, seemingly unaffected by the masers in nearby rungs.
However, the (12,6) rung -- like the (11,8) rung and also highlighted by a diamond in Figure~\ref{boltz} -- is not detected, suggestive of a statistically significant population deficit.
It is remarkable that the two most significant non-detections in our data set are for rungs above known masers.
If similar processes are at work in both the $K=6$ and $K=8$ ladders, then perhaps a population deficit is a general artifact of the upper-$J$ limit to the maser excitation mechanism.

In W51, Henkel et al.\ (2013) have also apparently detected the top of the maser ladder in several cases: in $K=6$ the physically associated $J=6$, 7, and 8 masers have no accompanying velocity component in (9,6); in $K=9$ the physically associated $J=9$, 10, and 11 masers have no accompanying maser in (12,9).
Although their velocity-drift method could not be applied to the $K=3$ ladder, it is intriguing to note that masers in $J=5$ and 6 are not accompanied by a maser in (7,3).
A determination of the thermal populations in the two rungs above the highest-lying maser rung in each W51 ladder -- that is, (13,9) \& (12,9), (10,6) \& (9,6), and (8,3) \& (7,3) -- would permit a valuable comparison to the possible deficit effects that we describe for NGC~7538.

One ladder-population result from Henkel et al.\ for para-ammonia in W51 is possible evidence against vibrational pumping.
The $K=4$ ladder has masers in (5,4) and (7,4) and a non-detection in (6,4).
If the $K=4$ masers in W51 can be physically associated using data from future observations, it will be the only known example of purely alternate-rung emission possible through simple rotational excitation.
That is, unlike the vibrational promotion suggested to explain the common-parity (10,8) and (9,8) masers in consecutive rungs presented in our current paper, the alternate-rung process of over-/under-population may be at work in the $K=4$ masers in W51.
In order to assess this possibility: (1) a physical association must be made between the (7,4) and (5,4) masers, and (2) long-integration single-dish observations must be made of (8,4) and (6,4) in order to constrain the thermal populations in the non-masering rungs adjacent to (7,4) and (5,4).
In general, this discussion of rungs above maser levels is a new advance for studies in this field: in all other cases of ammonia masers in the Galaxy, the rung above the highest-lying maser rung has not yet been observed.

\subsection{Model for Molecular Gas in IRS 1\label{model}}

\begin{figure}
\centering
\includegraphics[angle=000,scale=0.40]{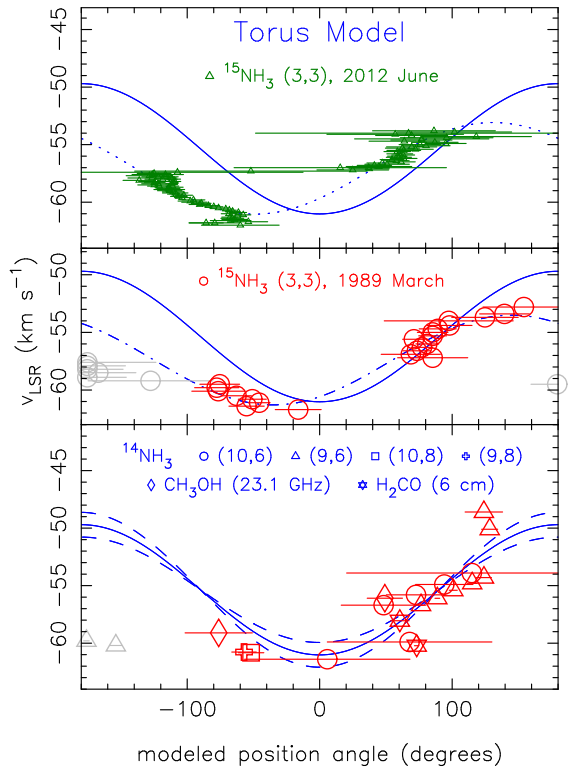}
\caption{
Computed position angles and measured velocities for interferometric observations of masers in NGC~7538 IRS~1.
The solid curve in each panel is the torus model of Surcis et al.\ (2011) described in Paper~III and Section~\ref{model}.
The dashed lines in the bottom panel are the uncertainty in this model.
In addition to the (10,6) and (9,6) masers plotted in Paper~III, to the bottom panel have been added the (10,8) and (9,8) masers from the current paper, as well as masers of CH$_3$OH and H$_2$CO calculated using archival data.
For the historic and current observations of $^{15}$NH$_3$ in the middle and top panels, respectively, are shown the best fits to each epoch (dot-dash for 1989, dotted for 2012).
The difference in best-fit position angle between 1989 and 2012 is discussed in Section~\ref{model} in terms of precession of the torus.
In all panels, the horizontal error bars are the 1-$\sigma$ uncertainty in calculated azimuthal position angle within the putative torus (the large uncertainties of the 2012 $^{15}$NH$_3$ masers near $v_{\rm LSR}=-57.3$~km/s are due to the small differences between their angular coordinates and those of the peak of continuum emission assumed to be the center of the rotation).
As discussed in Section~\ref{model}, the $^{15}$NH$_3$ masers with position angles near $180\degr$ do not fit the model (the gray circles in the middle panel), nor do the (9,6) masers near $180\degr$ (the gray triangles in the bottom panel, see Paper~III).
\label{torus}}
\end{figure}

Surcis et al.\ (2011) developed a model of a molecular torus in IRS~1 in order to explain the positions and velocities of 6.7-GHz methanol masers in the context of the mid-infrared disk described by De Buizer \& Minier (2005).
In the model, the velocity of the gas does not depend on the distance from the center of rotation and so the observed Doppler velocity depends only on the azimuthal location of the maser gas in the flow.
The model is computed for a torus with orbital inclination $i = 32\degr$ and an axis of rotation projected on the sky $\psi=40\degr$ west of north.
Thus, the sky location -- in offsets from the continuum peak $(\Delta\alpha_j,\Delta\delta_j)$ -- of the $j$th maser spot is converted into an azimuthal position angle $\phi_j$ in the torus by
\begin{eqnarray*}
	x_j &=& (\Delta\delta_j/\cos\psi - \Delta\alpha_j/\sin\psi)/(\cot\psi+\tan\psi) \\
	y_j &=& \cos{i}\ (\Delta\delta_j/\sin\psi + \Delta\alpha_j/\cos\psi)/(\cot\psi+\tan\psi) \\
	\phi_j &=& \tan^{-1}\left(\frac{y_j}{x_j}\right) + \phi_0 \ ,
\end{eqnarray*}
for which a value of $\phi_0\approx40\degr$ is the choice of fiducial for which the most negative modeled velocity lies at $\phi=0\degr$.
To our computed position angles $(\Delta\alpha_j,\Delta\delta_j)\rightarrow\phi_j$ of the masers is fitted the sinusoidal velocity pattern $v(\phi)$ expected from rotation.
\begin{eqnarray*}
	v(\phi) &=& v_{\rm rot}\,\cos(\phi - \phi_0') + v_{\rm sys}\ ,
\end{eqnarray*}
where $v_{\rm rot}$ is the rotation of the torus and $v_{\rm sys}$ is the secular translation of the rotating system.
In Paper~III, we applied the model to the $^{14}$NH$_3$ (10,6) and (9,6) masers.
The positions and velocities of the (10,6) and (9,6) masers are well fit by $v_{\rm rot} = -5.7\pm1.1$~km/s, $v_{\rm sys} = -55.4$~km/s, and $\phi_0'=0\degr$.
This fit is shown as a solid curve in all three panels of Figure~\ref{torus}.

The $^{14}$NH$_3$ (10,8) and (9,8) masers presented in the current paper also agree well with the model, as shown in the bottom panel of Figure~\ref{torus}.
The para-ammonia masers are unique among $^{14}$NH$_3$ masers in that they lie on the near side of the torus; the (10,6) maser with a similar velocity has a different position and the two (9,6) masers near position angle $-170\degr$ are excluded from the model (see Paper~III).
Furthermore, we have analyzed archival data for 6-cm H$_2$CO (Hoffman et al.\ 2003) and 23.1-GHz CH$_3$OH (Galv\'an-Madrid et al.\ 2010; Paper~I) and find agreement with the torus model for these species, as well (Fig.~\ref{torus}).
The 23.1-GHz methanol masers agree well with the model, including a location on the near side of the torus where Surcis et al.\ place the 6.7-GHz methanol groups B and E.
Although the H$_2$CO masers also show marginal agreement with the model, the agreement may be coincidental since the physical conditions of the formaldehyde maser environment are not expected within the disk/torus (formaldehyde requires lower densities and temperatures than ammonia, see Hoffman et al.\ 2003): the formaldehyde masers are suggested to lie in the outflow/cavity region of IRS~1 in order to be pumped by shocks/electrons and to attribute variability to a precessing jet (see Araya et al.\ 2007 and references therein).

We have also analyzed the $^{15}$NH$_3$ (3,3) masers in the context of the torus model.
Using the published positions and velocities from Gaume et al.\ (1991), we have computed the torus location of the masers for their epoch of observation in 1989 March.
As with the long-lived $^{14}$NH$_3$ (9,6) masers described in Paper~III, we find a group of $^{15}$NH$_3$ (3,3) masers near $v_{\rm LSR} = -60$~km/s with position angles that are not well modeled in the torus.
These masers are represented by the gray markers in the bottom and middle panels of Figure~\ref{torus}.
Excluding these masers from the fit, we find $v_{\rm rot}=-3.9\pm0.3$~km/s, $v_{\rm sys}=-57.4\pm0.2$~km/s, and $\phi_0'=-35.1\pm4.6\degr$.
This fit is shown as the dot-dash curve in the middle panel of Figure~\ref{torus}.
We also computed and fitted the torus model to the current epoch of $^{15}$NH$_3$ (3,3) observations from 2012 June, finding $v_{\rm rot}=-4.0\pm0.2$~km/s, $v_{\rm sys}=-57.08\pm0.07$~km/s, and $\phi_0'=-49.1\pm1.6\degr$, as shown as the dotted curve in the top panel of Figure~\ref{torus}.
The good agreement of $v_{\rm rot}$ and $v_{\rm sys}$ between the two epochs indicates that the azimuthal and translational dynamics of the $^{15}$NH$_3$ maser gas in the torus have been stable for the past 23 years, but that the plane of the torus has precessed $14.0\pm6.2\degr$, at an average rate of $0.6$~degrees per year, in the same sense as the rotation.\footnote{Note that motion due to rotation of the masers in a non-precessing torus would yield the same fitted value of $\phi_0'$ at every epoch.}
Kraus et al.\ (2006) have modeled a precessing jet in IRS~1 with a rate of 1.2 degrees per year, approximately twice our value.
For these rates, the precession periods are hundreds of years, presumably unrelated to the so-called long-term ($\approx 15$~year) periodic variability of masers in the region (Lekht et al.\ 2003, 2004; Hoffman et al.\ 2007).

The $^{15}$NH$_3$ masers have not shown significant intensity variability since their discovery in 1983 (Mauersberger, Wilson, \& Henkel 1986a; Johnston et al.\ 1989; Gaume et al.\ 1991).
Despite our identification of a new maser emission site (Fig.~\ref{15NH3}, \ref{gradient}) and our proposed rotational and precessional motion (Fig.~\ref{torus}), we find no significant difference between the integrated emission profile in 2012 June and any of those previously reported.
We do not track individual maser spots between epochs and so cannot constrain the lifetime of individual features.
(Nevertheless, we and Henkel et al.\ have observed several new masers to appear on $\sim25$-year timescales.)
Thus, the fate of the $^{15}$NH$_3$ (3,3) masers observed with $v_{\rm LSR} \approx -60$~km/s in 1989 March that were excluded from the torus fitting (the gray symbols in the middle panel of Fig.~\ref{torus}) is not known; since they are not observed in 2012 June, the gas at those locations may no longer be emitting beamed radiation on the line of sight or the emitting gas may have become entrained in the torus flow.

\subsection{Position Registration\label{registration}}

It is important to examine the method by which the association of the (10,8) and (9,8) masers is determined.
In particular, we find that the sub-arcsecond angular resolution afforded by interferometric observations is essential for including and excluding maser associations for pump modeling.
In Paper~III, we found fifteen (10,6) and (9,6) maser spots within 0\farcs5, only a few of which were coincident in position.
In contrast, in the literature it has been typical to group masers within $\sim 1$ arcsecond and having similar velocities into ``families'' (Wilson, Johnston, and Henkel, 1990; Henkel et al.\ 2013).
If IRS~1 were analyzed as single-dish data in this manner, a putative family of (10,6), (9,6), (10,8), and (9,8) masers would be associated with $v_{\rm LSR} = -60\,{\rm km}\,{\rm s}^{-1}$.
Of course, subsequent interferometric imaging has shown that this four-maser family exhibits the following disparate properties:
(1) only two of the masers -- (10,8) and (9,8) -- lie at the same position,
(2) only three of the masers are associated with a modeled torus (Fig.~\ref{torus}), and
(3) a separate class of (9,6) maser is suggested to exist as a subset of the complete sample in IRS~1.

With appropriate caution, then, we consider the studies of W51, which have produced most of the observational and theoretical development of nonmetastable ammonia masers.
Compared with NGC~7538, the interpretation of data for W51 is more susceptible to blending and false association for two reasons: W51 is twice as distant as NGC~7538 and the existing interferometric observations of W51 have 10-fold coarser angular resolution than used in Papers III and IV.
One possible false association is apparent: Henkel et al.\ observe four different velocity components for the (9,6) masers, one of which is categorized into a family with (5,3), (6,3), and (5,4) -- but the (9,6) maser is the only member of the family with pronounced intensity variability.
Considering the observed disparate behavior, common velocity alone is likely not sufficient evidence for associating masers physically.
However, common variability (such as the velocity drift observed by Henkel et al.\ for some families) appears to be a robust criterion for physical association.
Nevertheless, both Walsh et al.\ (2007) and Henkel et al.\ (2013) stress the importance of interferometric follow-up observations to single-dish surveys, and caution must be exercised when using spectrally- or angularly-blended data to constrain pumping models.

Throughout this series of papers, we have not compared maser positions from different observations by comparing absolute sky coordinates, but rather by comparing the following quantity: the relative separation between the maser emission and the peak of the continuum emission from IRS~1 in that same observation.
Because the continuum emission and the maser emission are detected simultaneously in the same band, the certainty in their relative separation is limited only by signal-to-noise (to $\sim 10$~mas), not by interferometer phase calibration.
Using the comprehensive study of the morphology of the continuum emission by Sandell et al.\ (2009), we register the continuum peaks from the different frequencies in order to determine maser positions.
Since many of the masers in question have frequencies $\sim 20$~GHz, only small morphology corrections are necessary and the registration is likely robust; one exception is the H$_2$CO masers at 4.8~GHz for which 200-mas uncertainties make difficult their inclusion in the torus model.
In general, compared with the continuum registration method used in our papers, the alternative registration method of comparing absolute coordinates has the potential for identifying false associations among masers and continuum features, as noted by Gaume et al.\ (1991).
For example, at least two different models of disks of 6.7-GHz methanol masers have been associated with the same feature in the image of the 22-GHz continuum emission (see De Buizer \& Minier 2005, and references therein).
The results from this paper and from Paper~III indicate that proper registration is crucial for modeling the molecular environment and maser excitation.

\section{Conclusion}

We present the discovery of para-ammonia masers in NGC~7538.
Interferometric observations of nonthermal emission from the $(J,K)=$(10,8) and (9,8) states
strongly indicate that the two masers arise in the same volume of gas.
Along with a non-detection of a maser in the (11,8) state, an excitation scheme involving radiative transitions out of the vibrational ground state may be constrained with these data.
Based on non-detections in the (11,6) and (12,6) states, we argue for similar constraints concerning excitations of ortho-ammonia masers out of the $K=6$ ground-state ladder (to either $K=0$ or to the $K=6$ vibrational excited state).

Because of the small angular separation (0\farcs5) among the ammonia masers in IRS~1, interferometric observations are required in order to describe fully the emission region.
The newly-discovered (10,8) and (9,8) masers fit well into an existing torus model of IRS~1.
With the advent of wider frequency coverage on interferometric arrays, modern instrumentation is increasingly suited to probing many rungs' emissions.
In general, the value of interferometric maser studies is two-fold:
first, with a pump model for the maser in hand, the selective physical conditions required for emission make each maser into a signpost for precisely measured values;
second, even without a pump model, the compact, bright, narrow maser lines allow unprecedented probes of the dynamics of the molecular environment.
The arguments presented in this paper have benefited from both of these factors.

The molecular torus model of Surcis et al.\ (2011) is used to explain the positions and velocities of several masers species, including ortho-ammonia, para-ammonia, methanol, and perhaps formaldehyde.
Even though the history of the modeling of NGC~7538 IRS~1 is one of many mutually exclusive disk/outflow models, the current torus model is shown to account for a wide range of maser phenomena.
The torus model has been adapted to include the precession apparent from interferometric $^{15}$NH$_3$ observations separated by $\sim25$~years.
The possibility of a single torus accounting for the many rare masers in IRS~1 is a valuable constraint on the study of 6-cm H$_2$CO masers and 23.1-GHz CH$_3$OH masers, for which pump mechanisms are not yet understood.
Furthermore, in cases where a single, common emitting volume is not required of different masers (unlike the para-ammonia masers in this paper), the ongoing search for positional coincidences among maser species may be suitably replaced by a search for entrainment at different positions in the same torus.
In order for the torus model to benefit the study of the rare masers in IRS~1, the model must be developed further to include the infrared illumination of the emitting volumes.

\acknowledgments

This work is supported by the Weaver Fund of Wittenberg University.
We are grateful to the GBT staff for technical support and for flexible scheduling of the telescope.
We are indebted to the editor and to an anonymous referee for helping to make this paper part of the literature.

{\it Facilities:} \facility{VLA ()}, \facility{GBT ()}

\begin{deluxetable*}{r r r | l r r r | l r@{.}l r }
\tablecolumns{11}
\tablewidth{0pt}
\tablecaption{Thermal Absorption GBT Observations and Results\label{absorption}}
\tablehead{
\multicolumn{3}{c}{Transition} & \multicolumn{4}{c}{Observational Parameters} & \multicolumn{4}{c}{Fitted Absorption Properties} \\
\multicolumn{1}{c}{State} & \multicolumn{1}{c}{$\nu_{\rm rest}$} & \multicolumn{1}{c}{$E_{\rm rung}$} & \multicolumn{1}{c}{Date} & \multicolumn{1}{c}{$\Delta\nu_{\rm BW}$} & \multicolumn{1}{c}{$N_{\rm chan}$} & \multicolumn{1}{c}{$t_{\rm int}$} & \multicolumn{1}{c}{$v_{\rm LSR}$} & \multicolumn{2}{c}{$\Delta{v}_{\rm FWHM}$} & \multicolumn{1}{c}{$T_{\rm line}$} \\
\multicolumn{1}{c}{$(J,K)$} & \multicolumn{1}{c}{(MHz)} & \multicolumn{1}{c}{(K)} & & \multicolumn{1}{c}{(MHz)} & & \multicolumn{1}{c}{(min)} & \multicolumn{1}{c}{(${\rm km}\,{\rm s}^{-1}$)} & \multicolumn{2}{c}{(${\rm km}\,{\rm s}^{-1}$)} & \multicolumn{1}{c}{(mK)}
}
\startdata
(15,12) & 18534.910 & 2644 & 2012 Nov 08 & 50.0 & 16384 & 16 & \tablenotemark{a} & \multicolumn{2}{c}{\tablenotemark{a}} & $>-$45 \\
(14,12) & 22353.910 & 2221 & \multicolumn{4}{l|}{\tablenotemark{b}} & \multicolumn{4}{c}{~} \\
(13,12) & 26654.850 & 1825 & \multicolumn{4}{l|}{\tablenotemark{b}} & \multicolumn{4}{c}{~} \\ \tableline
(16,9)  &  8824.180 & 3411 & 2012 Nov 07 & 50.0 & 32768 &  6 & \tablenotemark{a} & \multicolumn{2}{c}{\tablenotemark{a}} & $>-$35 \\
(15,9)  & 10754.530 & 2967 & 2012 Nov 07 & 50.0 & 32768 &  6 & \tablenotemark{a} & \multicolumn{2}{c}{\tablenotemark{a}} & $>-$40 \\
(14,9)  & 12951.050 & 2548 & 2013 May 04 & 50.0 & 16384 & 28 & \tablenotemark{a} & \multicolumn{2}{c}{\tablenotemark{a}} & $>-$15 \\
(13,9)  & 15412.490 & 2155 & \multicolumn{4}{l|}{\tablenotemark{b}} & \multicolumn{4}{c}{~} \\
(12,9)  & 18127.110 & 1789 & 2012 Nov 08 & 50.0 & 16384 & 16 & $-$59.1(5)  &  8&3(1) & $-$82(6) \\
(11,9)  & 21070.739 & 1449 & 2012 Mar 14 & 12.5 & 16384 & 32 & $-$59.5(5)  &  8&(1)  & $-$80(9)  \\
(10,9)  & 24205.287 & 1137 & 2012 Nov 08 & 50.0 & 16384 & 22 & $-$60.5(1)  &  5&5(3) & $-$176(4) \\
        &           &      &             &      &       &    & $-$55.9(2)  &  3&1(3) & $-$96(8)   \\ \tableline
(15,8)  &  9283.650 & 3052 & 2012 Nov 07 & 50.0 & 32768 & 10 & \tablenotemark{a} & \multicolumn{2}{c}{\tablenotemark{a}} & $>-$30 \\
(14,8)  & 11177.270 & 2635 & 2012 Nov 07 & 50.0 & 32768 & 10 & \tablenotemark{a} & \multicolumn{2}{c}{\tablenotemark{a}} & $>-$35 \\
(13,8)  & 13297.270 & 2242 & 2013 May 04 & 50.0 & 16384 & 28 & \tablenotemark{a} & \multicolumn{2}{c}{\tablenotemark{a}} & $>-$15 \\
(12,8)  & 15632.800 & 1878 & \multicolumn{4}{l|}{\tablenotemark{b}} & \multicolumn{4}{c}{~} \\
(11,8)  & 18162.250 & 1539 & 2012 Nov 08 & 50.0 & 16384 & 16 & \tablenotemark{a} & \multicolumn{2}{c}{\tablenotemark{a}} & $>-$90 \\
(10,8)  & 20852.527 & 1228 & 2012 Mar 14 & 12.5 & 16384 & 32 & $-$59.4(7)  & 10&(2)  & $-$58(9)  \\
(9,8)   & 23657.471 &  943 & 2012 Mar 14 & 12.5 & 16384 & 38 & $-$59.4(1)  &  7&3(3) & $-$134(6)  \\
        &           &      & 2012 Nov 08 & 50.0 & 16384 & 23 & $-$59.4(1)  &  7&8(3) & $-$132(6)  \\ \tableline
(14,6)  &  8766.970 & 2275 & 2012 Nov 07 & 50.0 & 32768 & 10 & \tablenotemark{a} & \multicolumn{2}{c}{\tablenotemark{a}} & $>-$25 \\
(13,6)  & 10426.890 & 2386 & 2012 Nov 07 & 50.0 & 32768 & 10 & \tablenotemark{a} & \multicolumn{2}{c}{\tablenotemark{a}} & $>-$35 \\
(12,6)  & 12251.330 & 2021 & 2013 May 04 & 50.0 & 16384 & 28 & \tablenotemark{a} & \multicolumn{2}{c}{\tablenotemark{a}} & $>-$20 \\
(11,6)  & 14224.650 & 1684 & 2013 May 04 & 50.0 & 16384 & 28 & $-$60.0(4)  &  5&2(9) & $-$13(1)  \\
(10,6)  & 16319.320 & 1375 & 2012 Jan 31 & \multicolumn{3}{l|}{\tablenotemark{c}} & \multicolumn{4}{c}{~} \\
(9,6)   & 18499.390 & 1092 & 2012 Nov 08 & \multicolumn{3}{l|}{\tablenotemark{c}} & \multicolumn{4}{c}{~} \\
(8,6)   & 20719.221 &  836 & 2012 Mar 14 & 12.5 & 16384 & 32 & $-$59.3(1)  &  6&4(2) & $-$292(9) \\
(7,6)   & 22924.940 &  608 & 2012 Mar 14 & 12.5 & 16384 & 38 & $-$59.44(5) &  7&2(1) & $-$328(5) \\ \tableline
(7,5)   & 20804.830 &  665 & 2012 Mar 14 & 12.5 & 16384 & 32 & $-$59.4(1)  &  5&5(3) & $-$226(12)  \\
(6,5)   & 22732.429 &  467 & 2012 Mar 14 & 12.5 & 16384 & 39 & $-$63.5(5)  &  2&7(7) & $-$54(9)  \\ 
        &           &      &             &      &       &    & $-$60.07(8) &  3&4(4) & $-$257(6) \\
        &           &      &             &      &       &    & $-$56.3(2)  &  2&7(3) & $-$93(7)  \\ \tableline
(12,3)  & 10836.130 & 2159 & 2012 Nov 07 & 50.0 & 32768 &  6 & \tablenotemark{a} & \multicolumn{2}{c}{\tablenotemark{a}} & $>-$45 \\
(11,3)  & 10536.180 & 1823 & 2012 Nov 07 & 50.0 & 32768 &  6 & \tablenotemark{a} & \multicolumn{2}{c}{\tablenotemark{a}} & $>-$70 \\
(10,3)  & 13296.340 & 1513 & 2013 May 04 & 50.0 & 16384 & 28 & \tablenotemark{a} & \multicolumn{2}{c}{\tablenotemark{a}} & $>-$15 \\
(9,3)   & 14376.820 & 1232 & \multicolumn{4}{l|}{\tablenotemark{b}} & \multicolumn{4}{c}{~} \\
(8,3)   & 16454.090 &  977 & 2012 Jan 31 &  1.0 &   256 & 17 & $-$59.7(2)  &  1&8(3) & $-$23(4)\tablenotemark{d} \\
(7,3)   & 18017.337 &  749 & 2012 Mar 14 & 12.5 & 16384 & 21 & $-$59.0(4)  &  8&6(9) & $-$110(9)  \\ 
(6,3)   & 19757.538 &  551 & 2012 Mar 14 & 12.5 & 16384 & 21 & $-$59.8(2)  &  8&3(4) & $-$236(9)  \\
(5,3)   & 21285.275 &  381 & 2012 Feb 26 & 12.5 & 16384 &  8 & $-$59.1(1)  &  7&3(3) & $-$476(16) \\
(4,3)   & 22688.312 &  237 & 2012 Feb 26 & 12.5 & 16384 &  3 & $-$66.7(3)  &  6&0(5) & $-$171(11) \\ 
        &           &      &             &      &       &    & $-$59.7(1)  &  6&8(5) & $-$379(7) \\
        &           &      &             &      &       &    & $-$51.6(3)  &  7&1(5) & $-$189(6) \\ \tableline
\enddata
\tablecomments{The number in parentheses is the uncertainty in the final digit.}
\tablenotetext{a}{~Line not detected. Limit for $T_{\rm line}$ is one standard deviation.}
\tablenotetext{b}{~This line was not part of the observational campaign.}
\tablenotetext{c}{~The maser emission observed for this transition is not accompanied by detectable absorption.}
\tablenotetext{d}{~$S_{\rm line}$ in units of mJy for $S_c=252$~mJy and $\eta=1$, represented by the triangle symbol in Figure~\ref{boltz}.}
\end{deluxetable*}


\begin{deluxetable*}{c r l c r r c | l r@{.}l r@{.}l r r@{.}l }
\tablecolumns{10}
\tablewidth{0pt}
\tablecaption{Maser Observations and Results\label{emission}}
\tablehead{
\multicolumn{7}{c}{Observational Parameters} & \multicolumn{8}{c}{Fitted Emission Properties} \\
\multicolumn{1}{c}{State} & \multicolumn{1}{c}{$\nu_{\rm rest}$} & \multicolumn{1}{c}{Date} & \multicolumn{1}{c}{Instr.} & \multicolumn{1}{c}{$\Delta\nu_{\rm BW}$} & \multicolumn{1}{c}{$N_{\rm chan}$} & \multicolumn{1}{c}{$t_{\rm int}$} & \multicolumn{1}{c}{$v_{\rm LSR}$} & \multicolumn{2}{c}{$\Delta{v}_{\rm FWHM}$} & \multicolumn{2}{c}{$S$} & \multicolumn{1}{c}{$\theta_{\rm maser}$} & \multicolumn{2}{c}{$T_B$} \\
\multicolumn{1}{c}{$(J,K)$} & \multicolumn{1}{c}{(MHz)} & & & \multicolumn{1}{c}{(MHz)} & & \multicolumn{1}{c}{(min)} & \multicolumn{1}{c}{(${\rm km}\,{\rm s}^{-1}$)} & \multicolumn{2}{c}{(${\rm km}\,{\rm s}^{-1}$)} & \multicolumn{2}{c}{(Jy)} & \multicolumn{1}{c}{(mas)} & \multicolumn{2}{c}{($10^4$~K)}
}
\startdata
(10,8) & 20852.527 & 2012 Mar 14 & GBT & 12.5 & 16384 & 32 & $-$60.82(5) & 1&1(1)  & 0&09(1) & \tablenotemark{a} & \multicolumn{2}{c}{\tablenotemark{a}}  \\
       &           & 2012 Jul 17 & VLA &  2.0 &   256 & 45 & $-$60.83(2) & 0&78(5) & 0&14(1) &     80 &   6&2 \\ \tableline
 (9,8) & 23657.471 & 2012 Mar 14 & GBT & 12.5 & 16384 & 38 & $-$60.78(2) & 0&69(6) & 0&09(1) & \tablenotemark{a} & \multicolumn{2}{c}{\tablenotemark{a}}  \\
       &           & 2012 Jul 17 & VLA &  2.0 &   256 & 45 & $-$60.72(2) & 0&57(6) & 0&15(1) &    100 &   2&7 \\
       &           & 2012 Nov 08 & GBT & 50.0 & 16384 & 24 & $-$60.77(2) & 0&60(4) & 0&13(1) & \tablenotemark{a} & \multicolumn{2}{c}{\tablenotemark{a}}  \\ \tableline
\enddata
\tablecomments{The number in parentheses is the uncertainty in the final digit. The fitted angular sizes of the masers have a precision of one significant digit, from which the brightness temperatures are calculated.}
\tablenotetext{a}{~Unresolved limit superseded by interferometric measurement.}
\end{deluxetable*}


\begin{thebibliography}{}
\bibitem[Araya et al.\ (2007)]{ara07} Araya, E., Hofner, P., Goss, W.\ M.\ 2007, 2007IAUS..242..110A, doi: 10.1017/S1743921307012653
\bibitem[Brown \& Cragg (1991)]{bro91} Brown, R.\ D., Cragg, D.\ M.\ 1991, \apj, 378, 445
\bibitem[De Buizer \& Minier (2005)]{deb05} De Buizer, James M., Minier, Vincent 2005, \apjl, 628, L151
\bibitem[Galv\'an-Madrid et al.\ (2010)]{gal10} Galv\'an-Madrid, R., Montes, G., Ram\'{\i}rez, E.\ A., Kurtz, S., Araya, E., Hofner, P.\ 2010, \apj, 713, 423
\bibitem[Gaume et al.\ (1991)]{gau91} Gaume, R.\ A., Johnston, K.\ J., Nguyen, H.\ A., Wilson, T.\ L., Dickel, H.\ R., Goss, W.\ M., Wright, M.\ C.\ H.\ 1991, \apj, 376, 608
\bibitem[Gaume et al.\ (1993)]{gau93} Gaume, R.\ A., Johnston, K.\ J., Wilson, T.\ L.\ 1993, \apj, 417, 645
\bibitem[Henkel et al.\ (2013)]{hen13} Henkel, C., Wilson, T.\ L., Asiri, H., Mauersberger, R.\ 2013, \aap, 549, A90
\bibitem[Hoffman et al.\ (2003)]{H03} Hoffman, I.\ M., Goss, W.\ M., Palmer, P., Richards, A.\ M.\ S.\ 2003, \apj, 598, 1061
\bibitem[Hoffman et al.\ (2007)]{H07} Hoffman, I.\ M., Goss, W.\ M., Palmer, P.\ 2007, \apj, 654, 971
\bibitem[Hoffman \& Kim (2011a)]{hof11a} Hoffman, Ian M.\ \& Kim, Stella Seojin 2011a, \apjl, 739, L15 (Paper~I)
\bibitem[Hoffman (2012)]{hof12} Hoffman, Ian M.\ 2012, \apj, 759, 76 (Paper~III)
\bibitem[Johnston et al.\ (1989)]{joh89} Johnston, K.\ J., Stolovy, S.\ R., Wilson, T.\ L., Henkel, C., Mauersberger, R.\ 1989, \apjl, 343, L41
\bibitem[Kraus et al.\ (2006)]{kra06} Kraus, S., Balega, Y., Elitzur, M., Hofmann, K.-H., Preibisch, Th., Rosen, A., Schertl, D., Weigelt, G., Young, E.\ T.\ 2006, \aap, 455, 521
\bibitem[Lekht et al.\ (2003)]{lek03} Lekht, E.\ E., Munitsyn, V.\ A., Tolmachev, A.\ M.\ 2003 Astron.\ Rep.\ 47, 838
\bibitem[Lekht et al.\ (2004)]{lek04} Lekht, E.\ E., Munitsyn, V.\ A., Tolmachev, A.\ M.\ 2004 Astron.\ Rep.\ 48, 200
\bibitem[Madden et al.\ (1986)]{mad86} Madden, S.\ C., Irvine, W.\ M., Matthews, H.\ E., Brown, R.\ D., \& Godfrey, P.\ D.\ 1986, \apjl, 300, L79
\bibitem[Mauersberger et al.\ (1986)]{mau86a} Mauersberger, R., Wilson, T.\ L., \& Henkel, C.\ 1986a, \aap, 160, L13
\bibitem[Mauersberger et al.\ (1986)]{mau86b} Mauersberger, R., Henkel, C., Wilson, T.\ L., \& Walmsley, C.\ M.\ 1986b, \aap, 162, 199
\bibitem[Mauersberger et al.\ (1987)]{mau87} Mauersberger, R., Henkel, C., Wilson, T.\ L.\ 1987, \aap, 173, 352
\bibitem[Mauersberger et al.\ (1988)]{mau88} Mauersberger, R., Henkel, C., Wilson, T.\ L.\ 1988, \aap, 205, 235
\bibitem[Moscadelli et al.\ (2009)]{mos09} Moscadelli, L., Reid, M.\ J., Menten, K.\ M., Brunthaler, A., Zheng, X.\ W., Xu, Y.\ 2009, \apj, 693, 406
\bibitem[Pratap et al.\ (1991)]{pra91} Pratap, P., Menten, K.\ M., Reid, M.\ J., Moran, J.\ M., Walmsley, C.\ M.\ 1991, \apjl, 373, L13
\bibitem[Sandell et al.\ (2009)]{san09} Sandell, G., Goss, W.\ M., Wright, M., \& Corder, S.\ 2009, \apjl, 699, L31
\bibitem[Schilke et al.\ (1991)]{sch91} Schilke, P., Walmsley, C.\ M., Mauersberger, R.\ 1991, \aap, 247, 516
\bibitem[Surcis et al.\ (2011)]{sur11} Surcis, G., Vlemmings, W.\ H.\ T., Torres, R.\ M., van Langevelde, H.\ J., Hutawarakorn Kramer, B.\ 2011, \aap, 533, A47
\bibitem[Walsh et al.\ (2007)]{wal07} Walsh, A.\ J., Longmore, S.\ N., Thorwirth, S., Urquhart, J.\ S., Purcell, C.\ R.\ 2007, \mnras, 382, L35
\bibitem[Walsh et al.\ (2011)]{wal11} Walsh, A.\ J.\ et al.\ 2011, \mnras, 416, 1764
\bibitem[Willey et al.\ (1995)]{wil95} Willey, Daniel R., Southwick, Rhonda Jo, Ramadas, Krishnan, Rapela, Brian K., Neff, W.\ Allen 1995, \prl, 74, 5216
\bibitem[Wilson et al.\ (1990)]{wil90} Wilson, T.\ L., Johnston, K.\ J., Henkel, C.\ 1990, \aap, 229, L1
\end{thebibliography}
\end{document}